\documentclass[noinfoline,stslayout]{imsart}

\usepackage{amsfonts,amssymb,amsmath,amscd,amsthm,latexsym}
\usepackage{natbib}
\usepackage{graphicx}
\usepackage[]{units}

\renewcommand\b{($\mathfrak{b}$)}
\newcommand\f{($\mathfrak{f}$)}
\newcommand\oh{\nicefrac{1}{2}}
\newcommand\bx{\mathbf{x}}
\newcommand\bX{\mathbf{X}}
\newcommand\JLP{Jeffreys--Lindley's paradox}

\begin{document}

\begin{frontmatter}

\title{On the Jeffreys--Lindley's paradox\protect\thanksref{T1}}
\runtitle{On the Jeffreys--Lindley's paradox}
\runauthor{Christian P.~Robert}
\thankstext{T1}{Christian P. Robert, CEREMADE, Universit{\' e} Paris-Dauphine, 75775 Paris cedex 16, France
{\sf xian@ceremade.dauphine.fr} and Department of Statistics, University of Warwick, 
Coventry CV4 8UW, United Kingdom. Research partly supported by the Agence Nationale de la Recherche (ANR,
212, rue de Bercy 75012 Paris) through the 2012--2015 grant ANR-11-BS01-0010 ``Calibration'' and by an Institut 
Universitaire de France chair. The author is most sincerely grateful to the editorial team for its comments and
support towards improving the exposition in the paper.}

\begin{aug}
 \author{\snm{{\sc Christian P.~Robert}}}
 \affiliation{Universit{\'e} Paris-Dauphine, CEREMADE, University of Warwick, Department of Statistics, and CREST, Paris}
\end{aug}

\begin{abstract}
This paper discusses the dual interpretation of the Jeffreys--Lindley's paradox
associated with Bayesian posterior probabilities and Bayes factors, both as a
differentiation between frequentist and Bayesian statistics and as a pointer to
the difficulty of using improper priors while testing. We stress the
considerable impact of this paradox on the foundations of both classical and
Bayesian statistics. While assessing existing resolutions of the paradox, we
focus on a critical viewpoint of the paradox discussed by \cite{spanos:2013} in
the current journal.
\end{abstract}

\end{frontmatter}

\section{Introduction}

In the statistical literature, there is little debate as to whether or not testing statistical hypotheses is the
most controversial aspect of statistical inference, with at least three major competing
schools approaching the problem from different angles and often concluding with
opposite decisions. In this regard, Lindley's (1957) paradox may constitute the most
quoted instance of the opposition between the frequentist and Bayesian schools of
inference. Two recent reassessments of the paradox appeared in {\em Philosophy of
Science}, with Spanos (2013) and Sprenger (2013) diverging in their resolution of the
paradox, which prompted me to reconsider in turn this fundamental argument both in the
frequentist-Bayesian debate and in the derivation of (more) coherent testing procedures
within the Bayesian framework.

Let me first recall the setting of the paradox as exposed in Lindley (1957),
often called {\em the Jeffreys--Lindley's paradox} after Dennis Lindley pointed out the
facts were already exposed in Jeffreys (1939).  Given a sample of size $n$ from
a normal distribution $\mathcal{N}(\theta,\sigma^2)$ with known variance
$\sigma^2$, testing whether or not the null hypothesis $H_0:\,\theta=\theta_0$ 
on the mean holds (against the alternative $H_1:\,\theta\ne\theta_0$) may lead to opposite
conclusions depending on the statistical perspective adopted to conduct the test.  Namely,
summarising the dataset into the sufficient statistic
$$
\bar x_n \sim \mathcal{N}(\theta,\sigma^2/n)
$$
leads to the $t$ statistic $t_n=\sqrt{n}(\bar x_n - \theta_0)/\sigma$ which is distributed as 
a $\mathcal{N}(0,1)$ variable under the null hypothesis, allowing for the computation of the
$p$-value equal to
$$
p(t_n)=\mathbb{P}(|T_n|>|t_n|)=1-2\Phi(|t_n|)\,,
$$
where $\mathbb{P}(\cdot)$ is a generic notation for a probability computation and $p(\cdot)$
is the symbol used here for the $p$-value function.
Relying upon this $p$-value to determine (or decide) whether or not $H_0$ holds means examining its 
numerical value either in absolute terms (as suggested by Fisher) or with respect to a bound
(as suggested by Neyman and Pearson). A Bayesian approach to the hypothesis testing problem, 
as exposed in Jeffreys (\citeyear{jeffreys:1939}) relies on the ratio of evidences (or 
marginal likelihoods) also called the Bayes factor (see Berger, 1985 or 
Robert, 2001). When the prior distribution on the parameter $\theta$ happens to be
the normal prior, $\theta\sim\mathcal{N}(\theta_0, \sigma^2)$, the Bayes factor
is given by
\begin{equation}\label{eq:Linbf}
\mathfrak{B}_{01}(t_n) = (1+n)^{\oh}\,\exp\left(-n t_n^2/2[1+n]\right)\,,
\end{equation}
which measures the evidence brought by the data in favour of the null hypothesis
relative to the alternative hypothesis. A decision about which hypothesis to select
is then based on the numerical value of $\mathfrak{B}_{01}(t_n)$, the default boundary
between null and alternative being $\mathfrak{B}_{01}(t_n)=1$, since the data then brings
the same evidence in favour of both hypotheses.

The paradox exposed by Lindley (1957) is that, {\em for a fixed numerical value
of $t_n$ and for almost any choice of prior distribution on $\theta$, the Bayes factor
$\mathfrak{B}_{01}(t_n)$ goes to infinity with the sample size $n$ while the
$p$-value $p(t_n)$ remains constant in $n$.} In Lindley's (1957) words, ``we [can be]
95\%~confident [as frequentists] that $\theta\ne\theta_0$ but have 95\%~belief
[as Bayesians] that $\theta=\theta_0$" (p.187). This occurs for instance when
$t_n=1.96$ and $n=16,818$ assuming the prior weights equally both hypotheses.  (And
for $n=164$ if $H_0$ is ten times more likely than $H_1$.) \cite{sprenger:2013}
takes the example of a test for extra-sensory capacities (ESP) to oppose a $p$-value
of $0.003$ and a Bayes factor of $12$ in favour of the null. This divergence of
outcomes is called a ``paradox" since the same dataset almost certainly leads
to opposite conclusions and hence decisions when the sample size $n$ is large.
It led many commentators of the paradox to conclude that one approach or the other 
was ``wrong".

While divergences between different statistical theories of inference and their
numerical conclusions are to be expected, the surprising phenomenon that they persist when the
sample size grows to infinity explains for the long-term impact of this paradox
and the fact that it is still the focus of attention for statisticians and
philosophers of science alike.  Although the frequentist-Bayesian opposition
expressed by the paradox can be thoroughly explained, as detailed in Section
2.1,  the Jeffreys-Lindley's paradox also has foundational
consequences within the Bayesian framework that are detailed in this paper.  

Indeed, my personal apprehension of the Jeffreys--Lindley's paradox is that it points
at the poor (and even unacceptable) behaviour of improper prior distributions
when testing point-null hypotheses. An illustration is the resolution proposed in
\citealp{robert:1993b}, aimed at suppressing the impact of an arbitrary
normalising constant in improper priors. This perspective is exposed in Section 
\ref{sub:echalotte}, while some possible Bayesian resolutions are indicated in Section
\ref{sec:chal}, without engaging the reader into a technical foray that would not bring
further insights on the concepts at work behind the Jeffreys--Lindley's paradox. However, 
a large majority of quotes and comments found in the literature view the paradox as 
an irreconcilable divergence between the Bayesian and the frequentist resolutions of the point-null hypothesis
testing problem, blaming (at least) one of those approaches for the
discrepancy. This point of view is debated and criticised in Section \ref{sub:warOroses}
and to a larger extent in Section \ref{sec:spamos}. As the paper starts with an analysis of the opposition
between the $p$-value and the Bayes factor (see, e.g.,
\citealp{kass:wasserman:1996}), I want to stress as a preliminary remark stage that my
Bayesian approach follows the decision-theoretic perspective advocated by Berger (1985),
which means that hypothesis testing is conducted with the intent of a course of
action (depending on the selection or rejection of $H_0$) rather than for the
epistemic attempt of uncovering the ``truth", agreeing in this respect with
the position advocated in Sprenger (2013).\footnote{% added in reply to Referee #1
Due to this stance, I sometimes refer to the Bayes factor as a "testing procedure" or even
as a "test", meaning it is the central tool to conduct the test.}

The plan of this paper is as follows: it reviews the different perspectives on the
paradox in Section \ref{sec:p!p}, it analyses and rebuts the recent criticism of
\cite{spanos:2013} in Section \ref{sec:spamos}, it studies some Bayesian resolutions of
the paradox in Section \ref{sec:chal}, and it concludes in Section \ref{sec:rflX}.

\section{Setting the paradox referentials}\label{sec:p!p}

\subsection{Frequentist versus Bayesian interpretations}\label{sub:warOroses}

Let us recall that the classical view of the \JLP~is that the Bayes factor and
the $p$-value asymptotically (in the sample size) differ to the point of leading
to opposite conclusions (acceptance versus rejection of the null hypothesis $H_0$).

There is obviously no mathematical issue with the paradox---otherwise it would
have been readily dismissed---: as the quantities involved in the two
perspectives evaluate different objects using different measures: the
probability measure of an event over the sample space versus the probability
measure of an event over the parameter space, the former being conditional on
the parameter value and the later on the observation of the sample. Despite the
large literature on the topic, I would also argue that this is {\em not} a
statistical paradox based on the argument that observing a constant
value\footnote{As pointed out by Lindley (1957): ``5\%~in to-day's small sample
does not mean the same as 5\%~in to-morrow's large one" (p.189).} of $t_n$ as
$n$ increases is not of statistical interest: when $H_0$ is true, $t_n$ has a
limiting ${\mathcal N}(0,1)$ distribution, which means the corresponding
$p$-value has a limiting uniform distribution, while, when $H_0$ does not hold,
$t_n$ converges almost surely to $\infty$, in which case both the Bayes factor
and the $p$-value converge to $0$. This behaviour is completely in line with
the general result of the consistency of the Bayes factor in this setting,
which is all too often  overlooked in most commentaries on the
Jeffreys--Lindley's paradox. And the Neyman--Pearson (frequentist) approach to
testing suggests decreasing both the Type I and Type II error, hence also
decreasing the acceptance boundary for the $p$-value when $n$ increases (see,
e.g., \citealp{lehmann:casella:1988}).

There are arguably several reasons why the two approaches, Bayesian \b~and
frequentist \f, should not numerically agree, even asymptotically. Those reasons
all revolve around the central feature that the Bayesian perspective is the only
one that allows probabilistic conditioning on the observed value $x_\text{obs}$
and solely on that value: 
\begin{itemize}
\item[(a)] one approach \b~operates on the parameter space $\Theta$, the range of the
possible values of $\theta$ under the alternative, while the other \f~is
produced exclusively on the sample space $\mathcal{X}$ under the null. They are thus covering
incompatible events. The same opposition occurs between confidence \f~and credibility \b~when constructing
interval estimators (a point also made by Sprenger, 2013);
\item[(b)] one \f~relies solely on the point-null hypothesis $H_0$ and on the
sampling distribution it induces, making an evaluation like the $p$-value absolute,
while the other \b~opposes the null $H_0$ and
its model to a marginal version of the models corresponding to $H_1$ (in the sense
that those models are integrated over the parameter space $\Theta$ against a
specific prior distribution), which implies that the Bayes factor and the posterior probability
of $H_0$ are relative;
\item[(c)] reproducing what may be the most famous quote from Jeffreys
(\citeyear{jeffreys:1939}, Section 7.2) one \f~could reject ``a hypothesis that
may be true (...) because it has not predicted observable results that have not
occurred" (which means considering the event $\{X>x_\text{obs}\}$, say, as the
new focus of inference, rather than the sole observation $x_\text{obs}$), in
contrast with the other \b~ which manages to condition upon the observed value
$x_\text{obs}$. This implies, in particular, that the former \f~cannot agree with
the likelihood principle \citep{birnbaum:1962}, while the other is almost
uniformly\footnote{Although I cannot digress further in this direction, the use
of improper priors as discussed in the following section is sometimes related
to a violation of the likelihood principle since they use the model to a
further extent than through the likelihood function. However, that such priors also face
difficulties when used for testing hypotheses is not to be overinterpreted, as
the difficulty simply stems from their impropriety.} in agreement with it
\citep{berger:wolpert:1988}; 
\item[(d)] at least in the Fisherian version of the frequentist perspective,
one \f~resorts to an arbitrarily fixed bound $\alpha$ on the $p$-value, while the
other \b~mostly refers to a threshold set by the experimenter, rather than the default boundary 
probability of $\nicefrac{1}{2}$.
In the later case, unless a genuine loss function on the consequences of a wrong decision or an
unbalanced prior weighting vector are constructed, both hypotheses are weighted
equally and the boundary signifies that the data favours one hypothesis versus
the other in terms of maginal likelihood values. This default principle is equivalent
to using the reference $0$--$1$ loss \citep{berger:1985} and to adopting the
value $1$ as the pivotal value for the Bayes factor.
\end{itemize}
A consequent literature \citep[see, e.g.,][]{berger:sellke:1987} has since then
shown how divergent those two approaches could prove (to the point of being
asymptotically incompatible). Despite the fact that both approaches are
consistent in the sense mentioned above, most commentators on the paradox
conclude by blaming the $p$-value (always rejecting at a given $\alpha$ level
for $n$ large enough), or the Bayes factor (always accepting for a fixed
$p$-value for $n$ large enough), or both (see, e.g., Spanos, 2013, and
Sprenger, 2013). Others (see, e.g., \cite{gelman:etal:2013}) have chosen to
bypass the opposition by considering tools at the interface between both
approaches, like posterior predictive checks.

\subsection{Improper inputs for Bayes factors}\label{sub:echalotte}

While the gap between the frequentist and the Bayesian degrees of evidence was
the reason for Lindley (1957) mentioning a statistical paradox, an orthogonal
consequence of Jeffreys's (\citeyear{jeffreys:1939}) and Lindley's
(\citeyear{lindley:1957}) exhibiting this paradox is to highlight the genuine
difficulty in using improper priors\footnote{Improper priors are extensions of
the standard probability measures on the parameter space to infinite mass
positive measures in order to reach more procedures and to close the
inferential scope in several senses, see, e.g., Robert (2001).} in testing
settings: as stressed by \cite{lindley:1957}, ``the only assumption that will
be questioned is the assignment of a prior distribution of any type" (p.188).
This was also the argument made by both \cite{shafer:1982} and
\cite{degroot:1982} (see also \citealp{degroot:1973}) in their discussion of
the paradox.  Note that, as discussed in Robert et al.'s
(\citeyear{robert:chopin:rousseau:2009}, Section 6.4) reassessment of his book,
Jeffreys (1939) does not address the general problem of using improper priors
in testing, namely that the Bayes factor may be undefined due to the lack of
normalising constants in such priors \citep{berger:1985,robert:2001}. Instead,
he relies on ad-hoc if effective solutions when available and more generaly sketches a second
(and under-appreciated) type of (proper) Jeffreys's priors for testing
statistical hypotheses.

%\section{The paradox, paradoxes, or non-paradox}\label{sec:p!p}

This second (but not secondary) level of interpretation for the paradox shifts
the asymptotics from the sample size to a prior scale factor. If we remain
within the normal framework of Lindley (1957), with one observation
$x\sim\mathcal{N}(\theta,\sigma^2)$, considering a prior distribution of the
form $\theta\sim\mathcal{N}(\theta_0,n\sigma^2)$ under the alternative
hypothesis leads to a Bayes factor that is identical to \eqref{eq:Linbf} for
$t_n=x$.  In this perspective, $n$ is a prior scale factor, so that the prior
variance is $n$ times larger than the observation variance.\footnote{In terms
of de Finetti's imaginary observations, the prior corresponds to the
information brought by $n$ less {\em imaginary observations} than the real
observations.} The interpretation of the phenomenon is then obviously different: {\em when the prior
scale $n$ goes to infinity, the Bayes factor goes to infinity no matter what
the value of the observation $x$ is}. (Note that both interpretations are
mathematically equivalent.) Under this new light, $n$ becomes what Lindley
(1957) calls ``a measure of lack of conviction about the null hypothesis"
(p.189), a sentence that I re-interpret as the prior (under $H_1$) getting more
and more diffuse as $n$ grows. However, I want to stress once again that nowhere
in Lindley's paper (nor in Jeffreys's book) is the difficulty with improper
priors spelled out clearly. 

In this (re-)interpretation of the \JLP, I consider that the phenomenon
exhibited therein is not paradoxical in the least: when the diffuseness of the
(alternative) prior (i.e., under $H_1$) increases, the only relevant piece of
information becomes that $\theta$ could be equal to $\theta_0$, to the extent
that it overwhelms any evidence to the contrary contained in the data.  For one
thing, and as put by Lindley (1957), ``the value $\theta_0$ is fundamentally
different from any value of $\theta \ne \theta_0$, however near $\theta_0$ it
might be" (p.189).\footnote{We will get return to this fundamental remark in
the discussion of \cite{spanos:2013} in the next section.} In addition, letting
$n$ grow to infinity means that the mass of the prior distribution in any fixed
neighbourhood of the null hypothesis and even in any set coherent with the data
at hand vanishes to zero.  There is therefore a deep coherence in the selection
of the null hypothesis $H_0$ in this case: being completely indecisive about
the alternative hypothesis means we could and should not choose this
alternative. It is not possible to pick the alternative hypothesis of an
undefined value of $\theta$ when opposed to the very special value $\theta_0$
if we want to be ``completely non-informative" about $\theta$ under $H_1$. This
analysis of the \JLP~is justifying (further) the prohibition of the use of
improper priors for testing point null hypotheses and selecting embedded models
(found for instance in DeGroot, 1982, Berger, 1985, and Robert, 2011).

Depending on one's perspective about the position of Bayesian statistics within
statistical theories of inference, one might see this as a strength or as a
weakness since Bayes factors and posterior probabilities do require a realistic
model under the alternative when $p$-values and Bayesian predictives do not.  A
logical reason for this requirement is that Bayesian inference need proceed
with the alternative model when the null is rejected. 
%added for referee #1
This Bayesian insight on the paradox therefore leads to the requirement to
handle testing statistical hypotheses under limited prior information and
within the paradigm. Solutions addressing this issue are discussed in Section
\ref{sec:chal}.  Prior to this, we provide and rebut the arguments given in
\cite{spanos:2013} that the Jeffreys-Lindley paradox actually highlights the
deficiencies of the Bayesian approach to testing.

\section{Don't be afraid...}\label{sec:spamos}

Under the rather provocative title {\em ``Who should be afraid of the
Jeffreys-Lindley paradox"\/},\footnote{Given the contents of the paper, the
author presumably intends Bayesian statistics or Bayesians as the recipients of
this question.} \cite{spanos:2013} offers his frequentist reassessment of the
paradox, arguing against both Bayesian and likelihood ratio approaches and in
favour of the postdata severity evaluation he and Mayo have both been
advocating since \citeyear{mayo:spanos:2004}.  Answering those criticisms was
the starting motivation for writing the current paper.

While I hope the reader is already familiar with the contents of Spanos
(2013), let me first recapitulate the main points made in this paper before
embarking onto a more detailed analysis of those arguments. Spanos (2013)
compares the frequentist (use of $p$-values), Bayesian (use of Bayes factors)
and ``likelihoodist" (use of likelihood ratios) approaches to statistical testing, with the
conclusion that the latter two ``give rise to highly fallacious results"
(p.75), while the $p$-value can be processed (or rescued) by the post-data severity analysis
of Mayo and Spanos (2004, defined below in Section 3.3)), escaping the \JLP~paradox. The paper insists on the
ability of this method to exhibit a certain degree $\gamma$ of discrepancy from
the null hypothesis, while Bayesian and likelihoodist methods cannot and do not
provide evidence for a particular alternative hypothesis (see, e.g., p.79).
Spanos (2013) concludes that the paradox ``has played an important role in
undermining the credibility of frequentist inference" (p.91) as being
"vulnerable to the fallacy of rejection" (p.91) but that the Bayes factor falls
prey to the "fallacy of acceptance".\footnote{The {\em fallacy of rejection} is
``(mis)interpreting reject $H_0$ (evidence against $H_0$) as evidence for a
particular $H_1$" (p.75), by which I understand for a specific value of the
parameter under the alternative hypothesis, and the {\em fallacy of acceptance}
is ``(mis)interpreting accept $H_0$ (no evidence against $H_0$) as evidence for
$H_0$" (p.75).}

In the following sections, I answer those criticisms by defending a
decision-based position on testing (Section \ref{sub:prosper}), refusing the
anti-Bayesian argument that the Jeffreys-Lindley's paradox only impact the
Bayesian resolutions (Section \ref{sub:Lindley.vs.Bayes}), before summarising
and considering Spanos' own solution based on severity (Section
\ref{sub:defence}) and on the need to calibrate the strength of the rejection
(Section \ref{sub:Lord_Foul}). My conclusion about those arguments is that,
while rejecting decisional premises, the severity perspective is inherently
depending upon a notion of significant difference (``substantive discrepancy",
p.88) or distance from the null.

\subsection{A proper decisional framework for testing}\label{sub:prosper}

The notion of {\em evidence} brought by the data in favour of or against an
hypothesis $H_0$ is never defined by Spanos (2013), even though it is
repeatedly mentioned throughout the paper.
%My experience is that the notion widely fluctuates according to its user,
%ranging from vague facts to specific mathematical constructs (see, e.g.,
%\citealp{skilling:2007a}). 
More importantly, there is no argument made therein as to what the specific
purpose of conducting a test (against, say, constructing a confidence interval)
is. \cite{spanos:2013} operates as though there
were an obvious truth ($H_0$ or $H_1$) and as though one and only one
statistical approach could reach it, despite the evidence to the contrary
represented by the consistency property of all three approaches in Lindley's (1957)
setting.\footnote{Ironically, the numerical example used in the paper (borrowed
from Stone, \citeyear{stone:1997}, also father to the marginalisation
paradoxes, see \citealp{dawid:stone:zidek:1973}) is the very same as Bayes's
billiard example (if with a larger value of $n$) and as Laplace's example on
births (with a similar value of $n$).} 

Indeed, what differentiates statistical tests from other aspects of statistical
inference like point estimation is that (a) there is a precise question being
asked about the statistical model under study, prior to observing the data, and
(b) the answer to this question will impact the subsequent actions of the
individual(s) who asked the question. 
Point (a) relates to Lindley's stress on the feature that the parameter value
$\theta_0$ is emminently special and quite different from any neighbouring
value. This value $\theta_0$ was selected for a reason and with a motive,
brought to the experimenter's attention by a theoretical construct, and this as
done prior to the observation stage rather than suggested from the data. From a
Bayesian viewpoint, this ultimate specificity implies that prior information is
available (to a certain degreee) as to why $\theta_0$ is a special value of the
parameter $\theta$.  
Point (b) is about assessing the consequences of the answer to the questions,
especially the wrong answer. Both from a frequentist and from a Bayesian
perspective, this assessment implies defining a loss or utility function that
quantifies the impact of a wrong answer and eventually determines the boundary
between acceptance and rejection.\footnote{This is the simplest type of loss
function: more advanced versions could include the case of a non-decision,
calling for more observations, as in \cite{berger:2003}.}

Spanos (2013) does not follow this decisional approach (which he considers as a
Trojan horse for validating Bayesian inference, see \citealp{spanos:2012}).
This is for instance the point made by the remark ``the problem does not lie
with the $p$-value or the accept/reject rules as such, but with how such
results are transformed into evidence for or against $H_0$ or a particular
alternative" (p.76). Thus, the error statistical approach he advocates (as
further discussed in Section \ref{sub:defence}) does not proceed from a
decisional step, even when handling an accept/reject outcome, but it instead
requires the introduction of a secondary $p$-value threshold, the {\em severity
evaluation}, coupled with a parameter value (or deviation) that represents a
significant distance from the null, a ``substantively significant" discrepancy
(p.88) in Spanos' terms. {\em In fine}, this interpretation of testing relies
on the use of an implicit loss function that sets which value of the parameter
is far and which is not. For instance, when Spanos (2013, p.75) states that
``there is nothing fallacious or paradoxical about a small $p$-value or a
rejection of the null, for a given significance level $\alpha$; when $n$ is
large enough, since a highly sensitive test is likely to pick up on tiny (in a
substantive sense) discrepancies from $H_0$", the ``substantive sense" can only
be gathered from a loss function. In connection with this notion of loss and of
distance from the null hypothesis, Spanos' side remark that ``what goes
wrong is that the Bayesian factor and the likelihoodist procedures use
Euclidean geometry to evaluate evidence for different hypotheses when in fact
the statistical testing space is curved" (p.90) carries little weight.  First,
it is mathematically incorrect given that the Bayes factor is invariant under
one-to-one reparameterisations of either the parameter or the sampling spaces,
hence impervious to the curvatures of those spaces\footnote{Spanos (2013, p.90
and p.91) uses the term ``statistical space" without a proper definition. It
can be either the parameter or the sample space since there is no decision
space in his axioms.} and to the choice of a specific geometry. Second, the
severity alternative put forward by Spanos in this paper rests upon the choice
of a divergence measure $d(\mathbf{X})$ which is most often Euclidean, while
the Bayesian and likelihood approaches rely on the likelihood function, which
does not rely on the choice of a (Euclidean or not) distance. 

\subsection{The paradox as an anti-Bayesian argument}\label{sub:Lindley.vs.Bayes}

\cite{spanos:2013}  argues that the Jeffreys-Lindley's paradox is demonstrating
against the Bayesian (and likelihood) resolutions of the problem by failing to
account for the large sample size.\footnote{His argument about the invariance
of the Bayes factor to $n$ (p.84) is found missing as the Bayes factor does
depend on $n$ as exhibited by $\mathfrak{B}(t_n)$ above.} As detailed in
Section \ref{sub:echalotte}, I do not disagree with this perspective to some
extent: I indeed consider that the most important lesson learned
from Lindley (1957) is that improper priors require special caution when
conducting point-null hypothesis testing. There is indeed little sense in
arguing in favour of a procedure that would always conclude by picking the
null, no matter what the value of the test statistics is. However, as pointed
out in Section \ref{sub:warOroses}, considering a fixed (in the sample size
$n$) value of the $t$ statistic has little meaning in an asymptotic
referential, i.e.~when $n$ increases to $\infty$. Either the $t$ statistic
converges in distribution to the standard normal distribution under the null
hypothesis $H_0$ or it diverges to infinity under the alternative $H_1$. This
is the reason why both the Bayesian and the likelihood ratio approaches are
consistent in this setting.\footnote{Somewhat in connection with this point, I
fail to understand why a Bayes factor would ``ignore the sampling distribution
(...) by invoking the likelihood principle" (p.90): the Bayes factor
incorporates the sampling distribution by integrating it out against the
associated prior under the alternative hypothesis. There is no invoking
involved and no likelihood principle at play in the construction of the
marginal likelihood, but solely an application of the rule of probability
calculus.} 

In an encompassing perspective about hypothesis testing, I do argue that the
Jeffreys-Lindley's paradox expresses foundational difficulties {\em for all} of
the three methodological threads discussed in Spanos (2013): when following
Fisher's approach, there is a theoretical and practical difficulty as to how
one should decrease the acceptance bound $\alpha=\alpha(n)$ on the $p$-value
when $n$ increases. This approach fails to provide a working and logical
principle from which this bound (or sequence of bounds) $\alpha(n)$ should be
chosen. For instance, the paper objects (p.78) that because ``of the large
sample size, it is often judicious to choose a small type I error, say
$\alpha=.003$" when this argument simply points at the arbitrariness of this
numerical value. In the specific setting of this example of Spanos (2013), it
is much worse, in that this bound could have been dictated by the data itself
since the observed $p$-value is equal to the nearby $.0027$. In addition, I
find the argument of consistency unconvincing in that case since both the Bayes
factor and the likelihood ratio tests are then consistent testing procedures. 

In the Neyman--Pearson referential, there is a fundamental difficulty in
finding a proper balance (or imbalance) between Type I and Type II errors,
since such balance is not provided by the theory, which settles for the
sub-optimal selection of a {\em fixed} Type I error. In addition, the whole
notion of {\em power}, while central to this referential, has arguable
foundations in that this is a {\em function} that inevitably depends on the
unknown parameter $\theta$. In particular, the power decreases to the Type I
error at the boundary between the null and the alternative hypotheses in the
parameter set.  For instance, referring to Spanos' (2013) arguments,
giving a meaning to the definition of severity (eqn.~(25), p.87) 
$$
\mathbb{P}(\bx;d(\bX)<d(\bx_0);\theta>\theta_1\text{ is false}) \,,
$$ 
where $x_0$ is the observable and $\bx$ should be $\bX$, seems impossible.
The third argument ``$\theta>\theta_1$ is false" that conditions this
probability statement makes no sense without a prior distribution on the parameter 
set.\footnote{An exchange with D. Mayo 
(2013, personal communication) led me to conclude that this probability is computed 
under the distribution of $\bX$ associated with the parameter $\theta_1$, where
$\theta_1$ is determined by the severity criterion, detailed in Section \ref{sub:defence}.} 
Even the corrected version
$$
\mathbb{P}_{\theta_1}(d(\bX)<d(\bx_0)) 
$$
depends on the choice of the particular alternative $\theta_1$ or has to be seen as
the power function which, like the risk function (see, e.g., Berger, 1985), prohibits most comparisons in a 
frequentist framework.

As discussed further in Section \ref{sub:echalotte}, if we abstain from
discussing the genuine difficulty in setting a joint prior distribution over
two distinct parameter spaces, following a standard Bayesian approach with a
flat (uniform) prior distribution on the binomial probability inferred about in
\cite{spanos:2013} leads to a Bayes factor of $8.115$ (p.80). Although this
quantity is larger than one, the calibration of the discrepancy from this
threshold constitutes the central difficulty in using Bayes factors towards
decision-making, Jeffreys's (1939) scale being highly formal despite being
often refered to \citep{kass:wasserman:1996}.

\subsection{Defending severity}\label{sub:defence}

\cite{spanos:2013} rebounds on the failures (or fallacies?) of all three main
approaches to address the difficulties with the Jeffreys--Lindley's paradox to
advocate his own criterion the ``postdata severity evaluation" introduced in
\cite{mayo:spanos:2004}.\footnote{\label{foot:2TO1}Section 6 starts with the
mathematically incorrect argument that, since we have observed $x_0$, in
connection with the null hypothesis $H_0:~\theta=\theta_0$, the sign of
$x_0-\theta_0$ ``indicates the relevant direction of departure from $H_0$".
First, random variables may take values both sides of $\theta_0$ for most
values of $\theta$. Second, the fact that one is testing $H_0$ against a
two-sided or a one-sided alternative hypothesis pertains to the motivation of
the test, not to the direction suggested by the data. The contentious
modification of the testing setting {\em once} the data is observed is a major
issue with Spanos' (2013) arguments, issue that we will discuss further below.} 
I recall that an ``hypothesis $H$ passes a severe test" if the data agrees with
$H$ and if it is highly probable that data not produced under $H$ agrees less
with $H$ (p.86).  While this sounds a reasonable desiderata, the notion of
severe tests has been advocated by Mayo and Spanos (2004) over the past years,
but it has not yet made a dent on the theory or on the practice of statistical
testing. As examplified by the paper (see, e.g., Table 1 on p.88 and the
discussion surrounding it), this solution requires more (user-based)
calibration than the regular $p$-value and it is thus bound to confuse
practitioners. Indeed, the severity evaluation as explained\footnote{Typos in
both the last line in p.87, which is mixing the standardised and the
non-standardised versions of the test statistic, and Table 1, which introduces
a superfluous minus sign, do not help in clarifying the issue.} 
in \cite{spanos:2013} implies defining for each departure from the null,
rewritten as $\theta_1=\theta_0+\gamma$, the probability that a dataset
associated with this parameter values ``accords less with $\theta>\theta_1$
than $x_0$ does" (p.87). (Note that, as discussed in footnote \ref{foot:2TO1},
the two-sided alternative has been turned {\em postdatum} into a one-sided
version. This is no more acceptable than stating that the date always supports
more the value at the maximum likelihood then at the null.) 

The notion of severity is therefore a mix of $p$-value and of Type II error
that is supposed to ``provide the `magnitude' of the warranted discrepancy from
the null" (p.88), i.e.~to decide about how close (in distance) to the null we
can get and still be able to discriminate the null from the alternative
hypotheses ``with very high probability" (p.86). The description found in Section 6
of Spanos (2013) implies a rejection of $H_0$ for the data at hand, based on the comparison
of the $p$-value with an acceptance bound, as in Fisher's perspective, followed by an
assessment of ``the largest discrepancy $\gamma$ from $H_0$ warranted by data $x_0$", which
derives a boundary parameter value $\theta_1$ from a severity level, $.9$ say. This
amounts to selecting a minimal power or maximal Type II error and to check for the corresponding
discrepancy to be ``substantively significant" (p.88), an assessment provided by the user and thus
defeating the original purpose of the approach.  The error statistical approach is therefore ineffective as
an operational tool, because of this double calibration that is required
from the user. Once more, as detailed in Spanos (2013), the value of this closest discrepancy
$\gamma$---which is thus a bound on where we can discriminate between $H_0$ and
$H_1$ for a given sample size $n$---does depend on another arbitrary tail
probability, the ``severity threshold", 
$$
\mathbb{P}_{\theta_1}\{d(\mathbf{X})\le d(x_0)\}
$$
introduced in eqn.~(25). This tail probability has to be chosen by the user without being more
intuitive or less subjective than the initial acceptance bound on the $p$-value.\footnote{When
considering the severity as a function of $\theta_1$, complement to a
probability cdf in $\theta_1$, its most natural interpretation would be
of a Bayesian nature, the bound being then a prior quantile. However, this solution is quite
unlikely to meet with the authors' approval.}
Furthermore, once the resulting discrepancy $\gamma$ is found, whether it is
far enough from the null is a matter of informed opinion since, as duly noted
by Spanos (2013), whether it is ``substantially significant (...) pertains to
the substantive subject matter" (p.88), implying once again some sort of loss
function or of prior information that the paper fails to
acknowledge.\footnote{While this is very much unlikely to be advocated either
by the author or by Bayesian statisticians, we note that, as a statistics, i.e.
a transform of the data, both the Bayes factor and the likelihood ratio could
be processed in exactly the same way to produce severity thresholds of their
own. See also \cite{johnson:rossel:2010} and \cite{johnson:2013} for the
related notion of uniformly most powerful Bayesian tests.} 

\subsection{Falling afoul of the fallacy of rejection}\label{sub:Lord_Foul}

In connection with the special meaning of the value $\theta_0$ and with the
argument of the fallacy of rejection, mentioned by Spanos (2013) as associated
with the $p$-value, several parts of his discussion of the Bayesian approach
state (see, e.g., p.81) that other values of $\theta$ are supported and even
{\em better} supported by the data than the null value $\theta_0$. This is a
surprising argument as (a) it pertains to the construction of Bayesian credible
intervals but not to testing and (b) it is a direct illustration of the
``fallacy of rejection" in that rejecting (or not) $H_0$ does not bring
evidence in favour of a particular value of $\theta$. While it is correct that
the observed data $\bx_0$ does ``favor certain values [of the parameter] more
strongly" (p.81) than $\theta_0$, those values are (a) driven by the data,
i.e.~will vary from one repetition of the statistical experiment to the next,
and (b) of no particular relevance for conducting a test, meaning that the
experimenter or the scientist behind the experiment had not expressed a
particular interest in those values before they were exposed by the data. The
tested value, $\theta_0=0.2$ say, is chosen prior to the experiment because it
has some special meaning for the problem or the theory at hand. The fact that
the likelihood and/or the posterior are/is larger in other values of $\theta$
does not constitute "conflicting evidence" (p.82) against the fact that the null
hypothesis holds. It simply reflects on the property that the likelihood
function is a random function of the parameter $\theta$, whose mode also varies
with the data and is almost surely not located at the true value of the
parameter. Since this is mathematical obvious, I find astounding that it
can be used as a logical argument against some statistical approaches to testing.

\section{Toward a resolution of the Bayesian version of the paradox}\label{sec:chal}

While the divergence between the frequentist and Bayesian answers is reflecting
upon the difference between the paradigms in terms of purpose and evaluation,
rather than condemning one of the approaches as implied by Spanos (2013),
the (Bayesian) debate about constructing limiting Bayes factors or posterior
probabilities that include improper prior modelling stands both open and
relevant. DeGroot's (1982) warning that ``diffuse prior distributions (...)
must be used with care" has now been impressed upon generations of students and
it is indeed a fair warning. There remains nonetheless a crucial need to
produce assessments of null hypotheses from a Bayesian perspective and under
limited prior information, once again without any incentive whatsoever to
mimic, reproduce or even come close to frequentist solutions like $p$-values.
%(I will therefore abstain from covering here the notion of {\em matching
%priors}, whose sole purpose is to bring frequentist and Bayesian coverages as
%close as possible, see e.g. \citealp{datta:mukerjee:2004}.)

\cite{robert:1993b} suggested selecting the prior weights of the two
hypotheses, $(\varrho_0, 1-\varrho_0)$ towards compensating for the increased
mass produced by the alternative hypothesis prior.\footnote{The compensation
cannot be of a probabilistic nature in that the overall mass of an improper
prior remains infinite for any weighting scheme.} While the solution therein
produced results that brought a numerical proximity with the (standard)
$p$-value, its construction is flawed from a measure-theoretic perspective
since the determination of the weights involves the value of the prior density
$\pi_1$ at the point-null value $\theta_0$, 
$$
\varrho_0 = (1-\varrho_0) \pi_1(\theta_0)\,,
$$
while a probability density is only defined almost everywhere. This difficulty
is shared by the Savage--Dickey paradox representing the Bayes factor solely in
terms of the prior density under the alternative hypothesis
\citep{robert:marin:2009}.\footnote{A solution to the above measure-theoretic
difficulties is to impose a version of $\pi_1$ that is continuous at $\theta_0$
so that $\pi_1(\theta_0)$ is uniquely defined. It however does not escape
controversy as it equates the values of two density functions under two
orthogonal measures, the Lebesgue measure and a Dirac measure.} I nonetheless
second this opinion that the degree of freedom represented by the prior weight
$\varrho_0$ in the Bayesian formalism should not be neglected to overcome the
difficulty in using improper priors.\footnote{Some will object at this choice
on Bayesian grounds as it implies that the prior does depend on the sample size
$n$.}

Among several available resolutions (see, e.g., Robert, 2001, Chapter 5),
a further step worth mentioning is Berger et al.'s
(\citeyear{berger:pericchi:varshavsky:1998}) partial validation of the use of
{\em identical} improper priors on the nuisance parameters, a notion already
entertained by Jeffreys (see the discussion in
\citealp{robert:chopin:rousseau:2009}, Section 6.3). While arguing about the
case of the``same" constant in both models as validating picking the ``same"
improper prior for both models has neither mathematical nor statistical
validation, relying on the same prior quite handily eliminates the major thorn
in the side of Bayesian testing of hypotheses. As demonstrated in
\cite{marin:robert:2007} and \cite{celeux:elanbari:marin:robert:2012}, it
allows in particular for the use of a partly improper $g$-prior in linear and
generalised linear models \citep{zellner:1986}.\footnote{Once again, choosing
$g=n$ should attract criticism from some Bayesian corners for being dependent
on the sample size, even though it boils down to picking an imaginary sample
\citep{smith:spiegelhalter:1982} size of $1$. See \cite{liang:etal:2008} for an
alternative approach setting an hyperprior on $g$.}

A last step towards the incorporation of improper priors within the Bayesian testing
paraphenalia is the recent investigation of the use of score functions $S(x,m)$ that 
extend the standard log score function associated with the Bayes factor:
$$
\log B_{12}(x) = \log m_1(x) - \log m_2(x) = S_0(x,m_1) - S_0(x,m_2)\,,
$$
where $m_i$ is the prior predictive associated with model $\mathfrak{M}_i$.
Indeed, there exists a whole family of proper scoring rules that are
independent from the normalising constant of the prior predictive
\citep{parry:dawid:lauritzen:2012} and can thus be used on improper priors as
well. For instance, Hyv\"arinen's (\citeyear{hyvarinen:2005}) score is one of
these scores. While the scores are delicate to calibrate, i.e.~the magnitude of
$S(x,m_1)-S(x,m_2)$ is not absolute, they provide a consistent method for
selecting models (Berger, 1985) and avoid the delicate issue of selecting
priors that differ for model selection and for regular inference (conditional
on the model). This is why Sprenger (2013) advises replacing the Bayes factor
with a logarithmic score, rewritten as
$$
\mathbb{E}^\pi\left[\mathbb{E}_\theta\{\log f(X|\theta)/ f(X|\theta_0)\}|x\right]\,,
$$
and compared with an acceptance bound. The Kullback--Leibler divergence used in
this score is utterly natural in terms of evaluating the impact of replacing
one distribution with the other. And, as stressed by Sprenger (2013), it does
not ``involve commitment to the truth or likelihood of $H_0$". The use of this
score however requires the choice of an acceptance bound, which calibration is
not provided by the theory.

\section{Reflections}\label{sec:rflX}

Even though I almost uniformly disagree with the presentation of the
Jeffreys--Lindley's paradox found in his paper, I am most grateful to Aris
Spanos for rekindling my interest in the paradox and inducing me to spell out
my thoughts on the topic in an organised manner. This paper has provided a
perspective on the foundations of Bayesian inference towards testing
statistical hypotheses, including the recovery of (some) improper priors for
this purpose, and on the reasons why the severity-based approach of
\cite{mayo:spanos:2004} fails as a convincing alternative to and criticism of the existing
branches of statistical hypothesis testing. I argued against the notion that
the discrepancy between frequentist and Bayesian procedures was a paradox, even
when occuring asymptotically. I also disputed the common argument that the Jeffreys--Lindley's 
paradox prohibits the use of improper priors, but instead called for the use of score
and predictive procedures in this context.

The appeal of great paradoxes\footnote{I use this term despite my reluctance to
call such phenomena ``paradoxes", since they correspond to neither logical
impossibilities nor to mathematical mistakes, but rather to contradictions in
reasoning or, in the current case, to attempts to bring two different paradigms
together.} is to address foundational issues in a field, either to reinforce
the arguments in favour of a given theory or, on the opposite, to cast serious
doubts on its validity. The fact that the Jeffreys--Lindley's paradox is still
discussed in papers (as exemplified by \citealp{spanos:2013} and
\citealp{sprenger:2013} in the current journal) and blogs, by statisticians and
non-statisticians alike, is a testimony to its impact on the debate about the
deepest foundations of statistical testing. The irrevocable opposition between
frequentist and Bayesian approaches to testing, but also the persistent impact
of the prior modelling in this case, are fundamental questions that have not
yet met with definitive answers. And they presumably never will for, as aptly
put by \cite{lad:2003}, ``the weight of Lindley's paradoxical result (...)
burdens proponents of the Bayesian practice". However, this is a burden with
highly positive features in that it paradoxically drives the field to higher
grounds, like the devising of novel decisional tools assessing deeper and
better the impact of a particular model choice, when compared with the Bayes
factor solution producing a unique number.\footnote{To conclude with a literary
quote, ``il faut imaginer Sisyphe heureux" \citep{camus:1942}.}

%\bibliographystyle{ims}
%\bibliography{biblio.bib}

\end{document}